\documentclass[aps,longbibliography,superscriptaddress,reprint]{revtex4-1}

\usepackage[utf8]{inputenc}
\usepackage{graphicx}
\usepackage{siunitx}
\usepackage{xcolor}
\usepackage{hyperref}
\usepackage{filecontents}

\usepackage{ulem}

\begin{document}

\title{Blue-detuned molecular magneto-optical trap schemes based on bayesian optimization}
		
\author{S. Xu}
\author{R. Li}
\affiliation{National Institute of Extremely-Weak Magnetic Field Infrastructure, Hangzhou, 310052,China}
\author{Y. Zhai}
\email{yueyangzhai@buaa.edu.cn}
\affiliation{Key Laboratory of Ultra-Weak Magnetic Field Measurement Technology, Ministry of Education,
School of Instrumentation and Optoelectronic Engineering, Beihang University, Beijing, 100191,
China}
\affiliation{Zhejiang Provincial Key Laboratory of Ultra-Weak Magnetic-Field Space and Applied Technology, Hangzhou Innovation Institute,
Beihang University, Hangzhou, 310051}
\affiliation{Hefei National Laboratory, Hefei, 230088, China}
\author{Y. Xia}
\email{yxia@phy.ecnu.edu.cn}
\affiliation{State Key Laboratory of Precision Spectroscopy, School of Physics and Electronic Science, East China Normal University, Shanghai 200241, China}
\author{M. Siercke}
\author{S. Ospelkaus}
\affiliation{Institut f\"ur Quantenoptik, Leibniz Universit\"at Hannover, 30167~Hannover, Germany}

\begin{abstract}
Direct laser cooling and trapping of molecules to temperature below Doppler limit and density
exceeding $10^8$ are challenging due to the sub-Doppler heating effects of molecular magneto-optical
trap (MOT). In our previous paper \cite{cite01}, we presented a general approach to engineering the sub-
Doppler force by tuning the AC stark shift with the addition of a blue detuned laser. Here, by
employing the Bayesian optimization method to optical Bloch equations, we have identified multiple 
blue-detuned MOT schemes for the CaF molecule. From the three-dimensional Monte-Carlo
simulation, we obtained a MOT temperature and density of \SI{14}{\micro K} and \SI{4.5e8}{cm^{-3}}, respectively.
Our findings present a potential avenue for directly loading molecular MOTs into conservative traps,
which can capitalize on the high density and low temperature of the MOT 
\end{abstract}

\maketitle
Recent experiments using molecular magneto-optical traps (MOT) have been found to suffer from the
sub-Doppler heating effects caused by the complex energy levels involved \cite{cite02,cite03}. 
These effects lead to higher temperature and lower density in the resulting molecular cloud \cite{cite04,cite05,cite06,cite07}. 
While blue-detuned molasses can effectively cool molecules to temperatures as low as a few \SI{}{\micro K} \cite{cite08,cite09,cite10,cite11,cite12,cite13}, 
they are not suited to compressing or increasing the density of the MOT. The successful implementation of a 
blue-detuned MOT using type II transition in Rb atoms is a good indication that this technique should work in molecules as well \cite{cite14}.
However, the close hyperfine splittings of molecule result in the interaction of one laser 
component with all energy levels, which makes it impossible to directly replicate the technique used in atomic MOT. 
Recently, the YO molecule’s distinctive energy level structure has enabled the successful implementation of 
a blue-detuned MOT \cite{cite15}, in which the phase-space density is increased by two orders of magnitude. In this paper, we
describe the utilization of Bayesian optimization to aid in the discovery of blue-detuned MOT schemes for CaF
molecules. We have successfully developed two separate MOT schemes for the $\sigma^-\sigma^-\sigma^+\sigma^-$ and $\sigma^+\sigma^+\sigma^+\sigma^-$ DC MOT configurations 
that are capable of providing cooling within 5 m/s and trapping within a 3 mm radius. We also discovered a scheme that utilizes 
only two laser components to achieve a competative MOT force, which will simplify the experiment significantly. Monte-Carlo simulation with
the MOT force results in a significant reduction in temperature and a corresponding increase in density, 
which can facilitate the transfer of a large number of molecules to conservative trap \cite{cite12,cite13,cite16,cite17,cite18,cite19} and thus 
enhance the effects of subsequent cooling such as evaporative or sympathetic cooling \cite{cite20,cite21,cite22,cite23,cite24}.
\begin{figure*}[htbp]
    \begin{center}
    \includegraphics[scale=0.6]{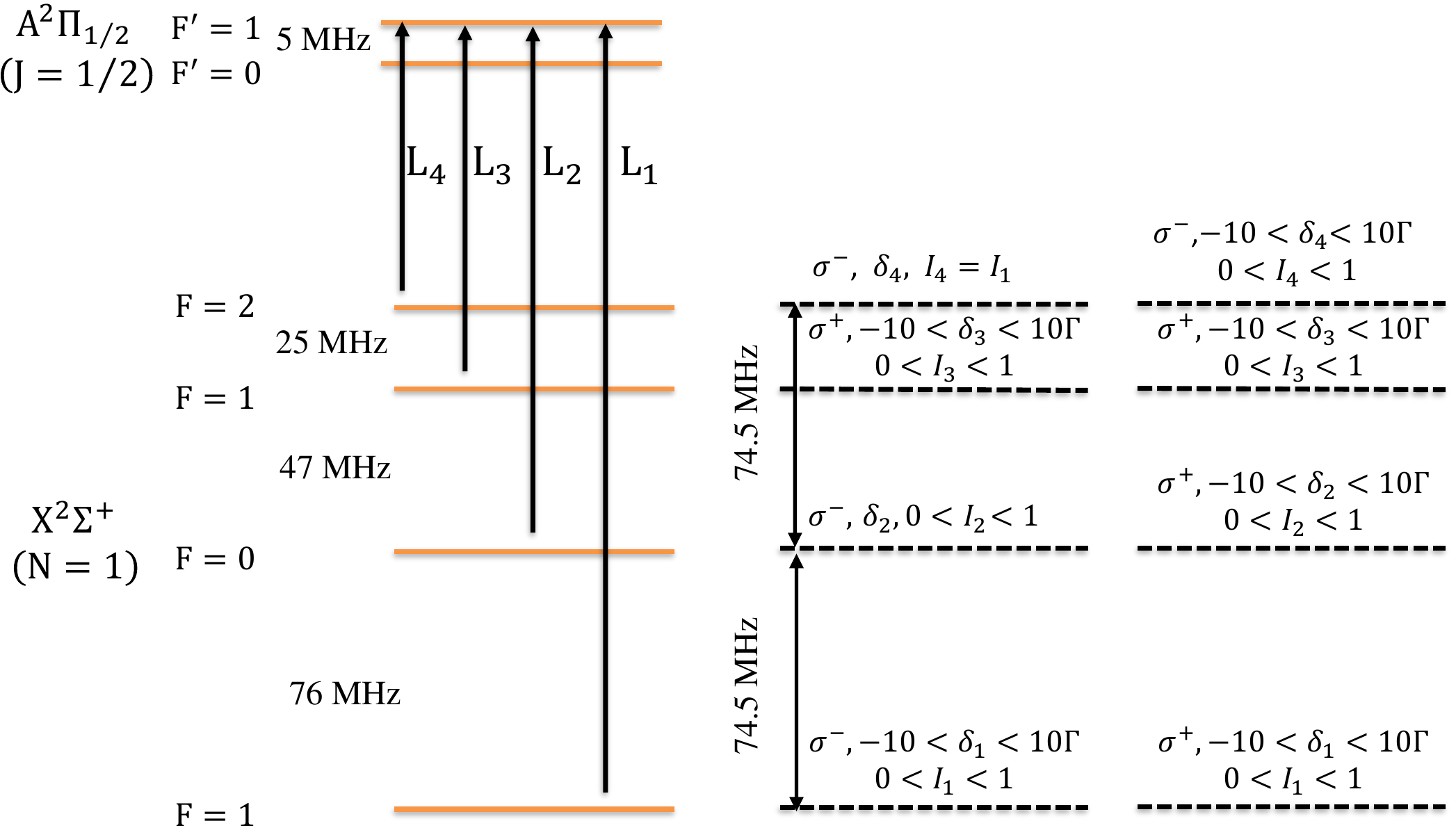}
    \caption{The energy levels of CaF and the MOT configurations require optimization.The detuning of each laser component
    can vary between $-10$ to $10\Gamma$, while the intensity ratio can range from 0 to 1. In the $\sigma^-\sigma^-\sigma^+\sigma^-$ configuration, 
    there are five free parameters, with $\delta_2$ and $\delta_4$ adjusting based on the modulation frequency of 74.5 MHz and the energy splittings 
    of the hyperfine states. On the other hand, the $\sigma^+\sigma^+\sigma^+\sigma^-$ configuration has eight free parameters, which can 
    be reduced to four if the laser intensity ratio is set equally}    
    \label{fig:1}
    \end{center}
\end{figure*}
\begin{figure}[b]
    \begin{center}
    \includegraphics[scale=0.6]{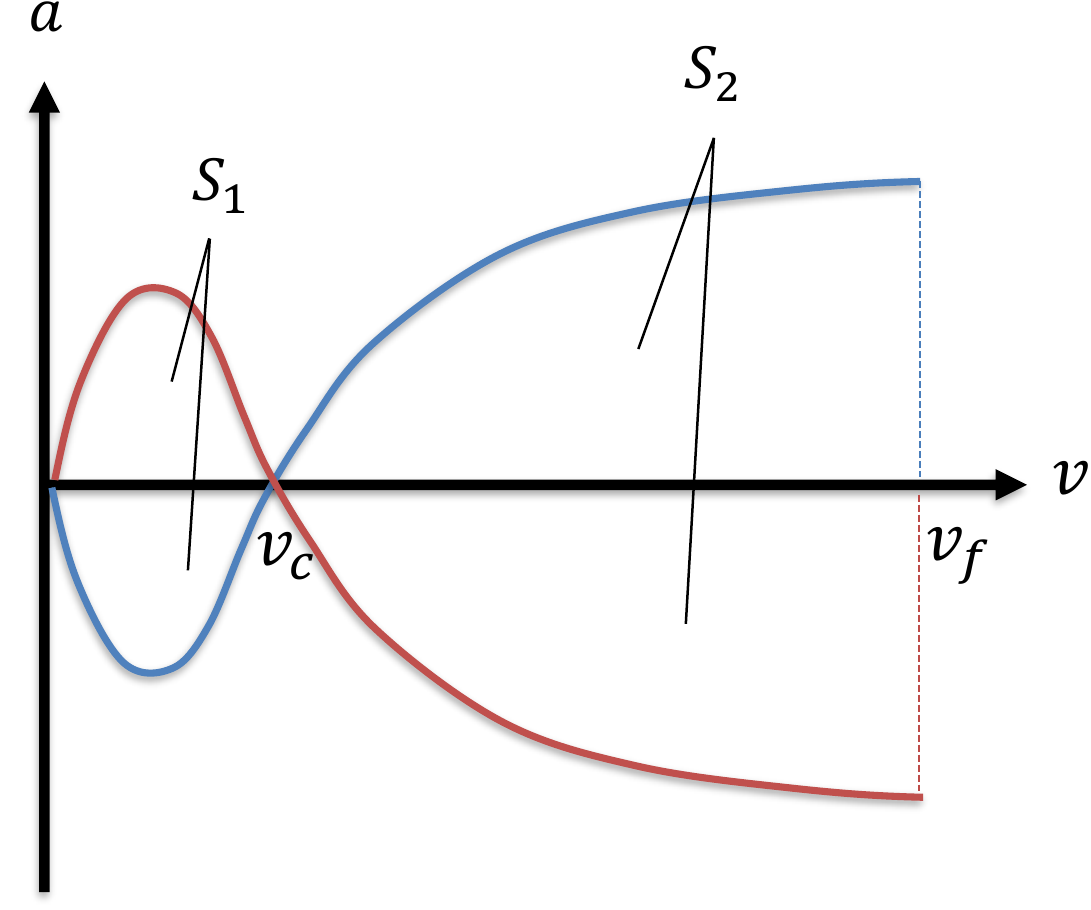}
    \caption{A red (blue) detuned MOT typically follows a cooling
    acceleration curve that involves sub-Doppler heating (cooling)
    in the low velocity range and Doppler cooling (heating) in
    the high velocity range. The critical velocity, denoted as $v_c$,
    marks the point at which the force sign changes. We typically
    calculate the velocity range $v_f = \Gamma/k$, where $\Gamma$ is the natural
    linewidth of the molecular transition and $k$ is the wave vector
    of the lasers}    
    \label{fig:2}
    \end{center}
\end{figure}

\begin{figure*}[htbp]
    \begin{center}
    \includegraphics[scale=0.5]{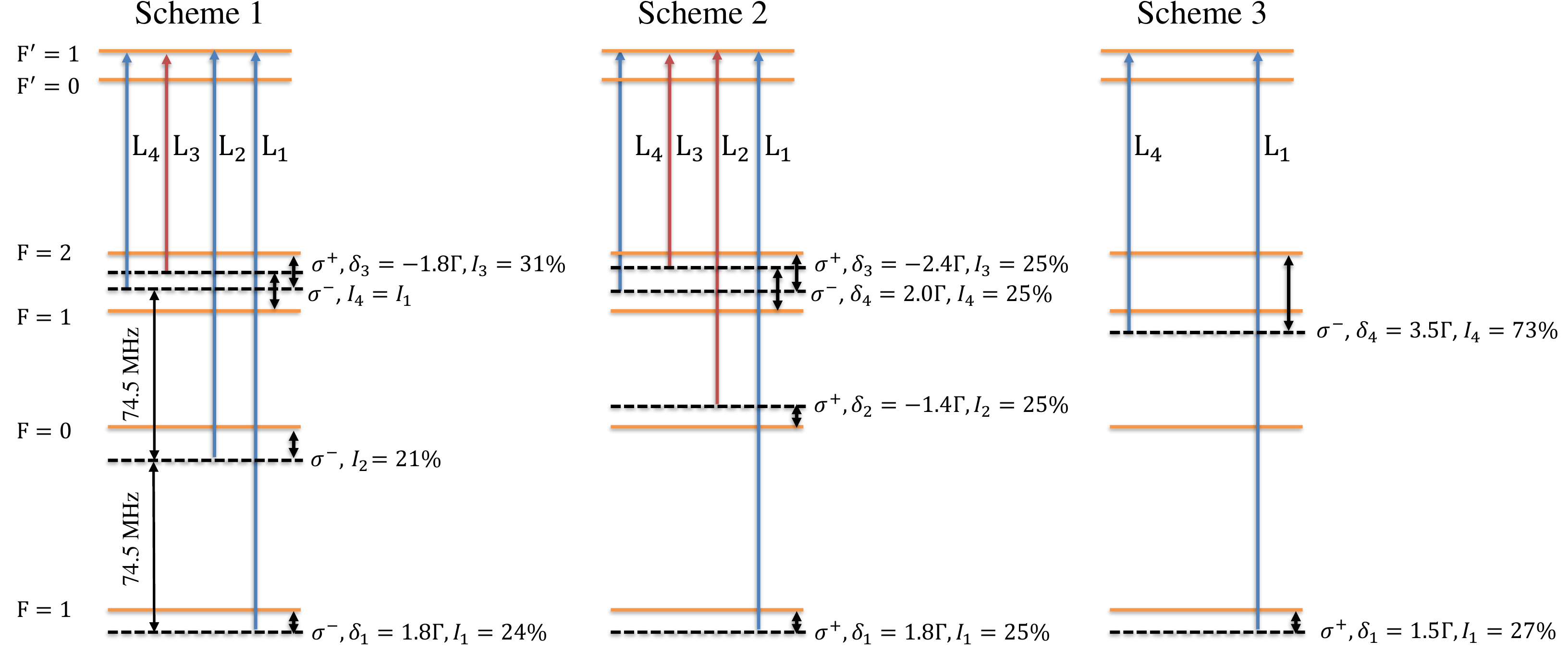}
    \caption{The blue-detuned MOT schemes for the CaF molecule discovered using Bayesian optimization. 
    Scheme 1 corresponds to the $\sigma^-\sigma^-\sigma^+\sigma^-$ configuration, while Scheme 2 
    corresponds to the $\sigma^+\sigma^+\sigma^+\sigma^-$ configuration. Scheme 3 represents the configuration with 
    two laser components. The blue (red) upward arrows
    indicate the blue (red) detuning relative to the target energy levels. The detunings and laser intensity ratios are labeled in the
    figure.}    
    \label{fig:3}
    \end{center}
\end{figure*}

\begin{figure*}[htbp]
    \begin{center}
    \includegraphics[scale=0.3]{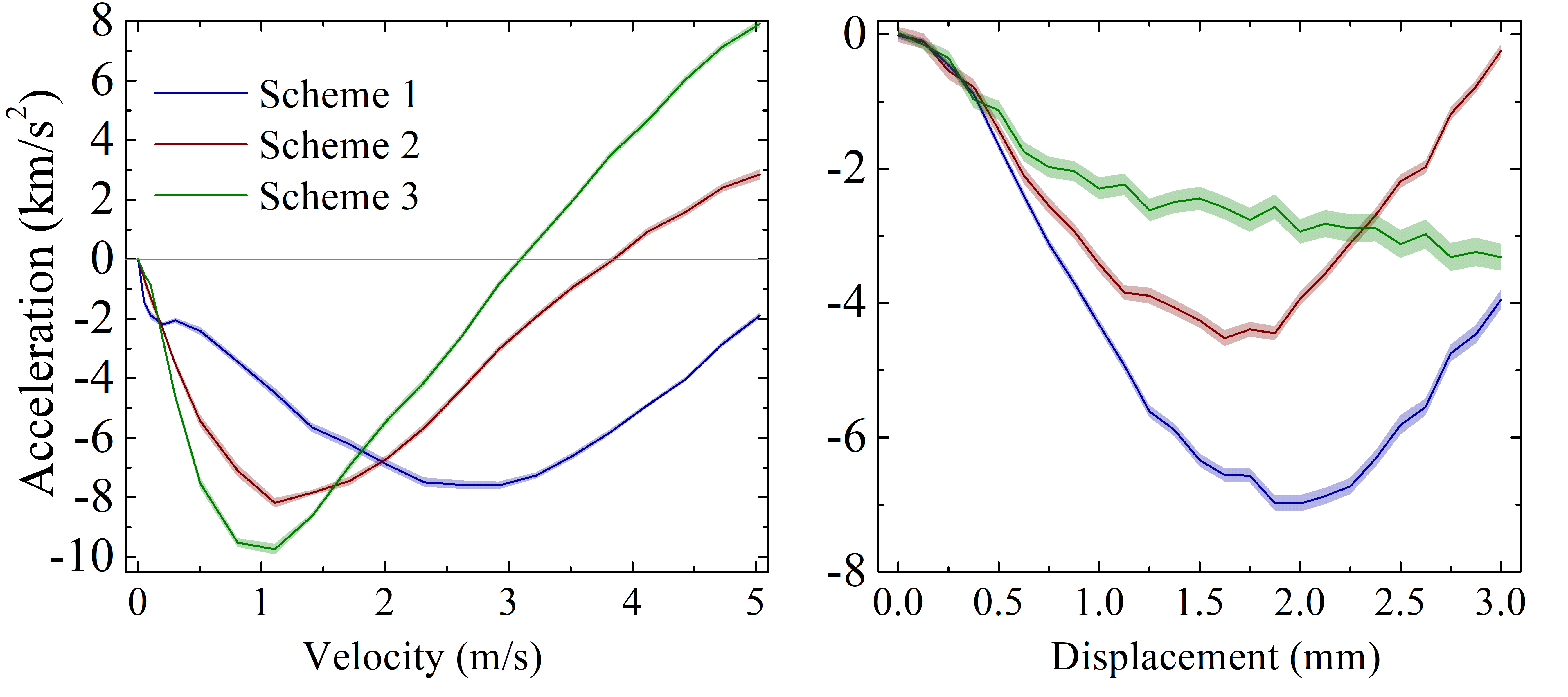}
    \caption{The acceleration curves for the optimal MOT schemes, showing the velocity-dependent force calculated up to $\Gamma/k$
    ($\sim$5 m/s) and the trapping force calculated up to 3 mm along the z direction. All schemes can effectively compress and cool
    the molecules within the MOT center. The shaded areas represent the 68\% confidence interval based on 100 and 1000 runs of
    OBEs for the cooling and trapping forces, respectively}    
    \label{fig:4}
    \end{center}
\end{figure*}
Fig. \ref{fig:1} shows the energy level structure of CaF molecule along with the parameters that require optimization.
The $\rm X^2\Sigma^+$ ground state consists of four hyperfine states, namely F = 1, 0, 1 and 2, with corresponding energy
splittings of 76 MHz, 47 MHz and 25 MHz, seperately. On the other hand, the $\rm A^{2}\Pi_{1/2}$ excited state has only
two hyperfine states, namely $\rm F^{'}=0$  and 1, and the energy difference between them is set to 5 MHz \cite{cite25}. The dashed
lines correspond to the four laser components that aim to target the four ground hyperfine states and the 
upward-pointing arrows indicate the transition to the excited $\rm F^{'}=1$ state. Starting from the bottom, the four laser
components are identified as $\rm L_1, L_2, L_3$ and  $\rm L_4$. Each of these components is associated with a specific polarization, 
detuning, and laser intensity ratio. To narrow down the parameter search space, we limit the polarization to 
$\sigma^-\sigma^-\sigma^+\sigma^-$ and $\sigma^+\sigma^+\sigma^+\sigma^-$($\sigma^-\sigma^-\sigma^+\sigma^-$ and $\sigma^+\sigma^+\sigma^+\sigma^-$ denote that 
$\rm L_1, L_2, L_3$ and  $\rm L_4$ have a polarization of $\sigma^-\sigma^-\sigma^+\sigma^-$ and $\sigma^+\sigma^+\sigma^+\sigma^-$, respectively),
since both of them are implemented in experiment \cite{cite04,cite05}. The laser beam has a total power of 100 mW and a radius
of 7.5 mm in all directions. The magnetic field gradient remains fixed at 15 G/cm. For $\sigma^-\sigma^-\sigma^+\sigma^-$ configuration,
we refer to the \cite{cite04} MOT scheme, in which a single laser modulated by an electric-optic modulator (EOM) generates 
the $\rm L_1, L_2$ and $\rm L_4$ laser components with a modulation frequency of 74.5 MHz. $\rm L_3$ has a distinct polarization
that is generated by an acousto-optic modulator (AOM). In total, we have five free parameters, as shown in the
picture, with their tuning ranges. We attempted two different schemes for the $\sigma^+\sigma^+\sigma^+\sigma^-$ configuration. The
first scheme involved restricting the laser intensity ratio to be equal, resulting in four different detunings as the
only free parameters. The second scheme included eight free parameters, which comprised the detunings and the 
laser intensity ratio.

Prior to beginning the optimization process, it is necessary to establish a clear goal for the program. This is
because we must strike a balance between two types of MOT forces: the cooling force, which depends on velocity, 
and the trapping force, which is dependent on space. We initially anticipate a cooling acceleration curve
similar to that shown in Fig. \ref{fig:2}. This curve exhibits two distinct indications of force for red (blue) detuning: 
sub-Doppler heating (cooling) in the low velocity range, and Doppler cooling (heating) in the high velocity range. 
In order to quantify the effectiveness of cooling acceleration with a single value, we employ two integrals: 
${S_1} = \sum_{v=0}^{v_c}a\Delta v$ and ${S_2} = \sum_{v=v_c}^{v_f}a\Delta v$. These integrals are used to 
calculate a metric for the quality of cooling acceleration, represented as $f_v = (10S_1 + S_2 )/v_f$. We
assign a weight of 10 to $S_1$ to highlight its significance, otherwise the program prefers to increasing the $S_2$ 
part instead of decreasing the $S_1$ part. The effectiveness of trapping acceleration $f_z$ is determined solely 
by the average value obtained within the range of 0 to 3 mm. We attempt to define the final goal for the 
program in two different ways. In the first approach, we optimize the program separately to attain the maximum 
cooling and trapping acceleration $f_{vm}$ and $f_{zm}$ . Then, we establish a final result of 
$R = (f_v /f_{vm} )+(f_z /f_{zm} )$ for the program to execute. Alternatively, in the second approach, we assign a 
dimensionless value of $R = f_v * f_z$ for the program to execute. After conducting multiple tests, we have determined 
that the second method yields better results and enables us to efficiently and consistently identify the
optimal group of parameters.

In order to capture the sub-Doppler force of the MOT, we run the optical Bloch equations (OBEs) \cite{cite03} for a sufficient 
amount of time until it reaches a quasi-steady state. We also select Julia as our simulation program to enhance the 
running speed. Bayesian optimization can be implemented using the Scikit-Optimize Python package, and can be called 
in a Julia environment using the PyCall package. Fig. \ref{fig:3} illustrates the blue-detuned MOT schemes we have found for 
CaF molecule. As we didn’t observe any significant improvement when using eight free parameters in the $\sigma^+\sigma^+\sigma^+\sigma^-$
configuration, for experimental convenience, we only display the results obtained with equal laser intensity. 
It is evident that in the $\sigma^-\sigma^-\sigma^+\sigma^-$ configuration, $\rm L_1$ is detuned by $1.8\Gamma$ with respect to 
the lower F = 1 to $\rm F^{'}=1$ transition and has an intensity ratio of 24\%. The detuning of $\rm L_2$ and $\rm L_4$ is 
dependent on the modulation frequency of 74.5 MHz and the energy splittings. $\rm L_2$ has an intensity ratio of 21\%, 
whereas the intensity of $\rm L_4$ is equal to that of $\rm L_1$. $\rm L_3$ is red detuned by $-1.8\Gamma$ relative to the upper F = 1 to 
$\rm F^{'}= 1$ transition and has an intensity of 31\%. The corresponding acceleration curve is illustrated in Fig. \ref{fig:4}, 
where we calculate the velocity-dependent force up to $\Gamma/k$ at the center of the MOT, and the space-dependent force along 
the z direction up to a distance of 3 mm, considering a velocity of 0.1 m/s. We observe that the cooling acceleration
in Scheme 1 remains negative and strong up to 5 m/s, with a slight bump at 0.2 m/s. Concurrently, the trapping acceleration 
exhibits a persistent strength within the computed range. During our search process, we notice that the program consistently
generates a strong MOT force even when the parameters fluctuate around the optimal values. This indicates the 
robutness of our results. Hence, for experimental convenience, it is possible to set the laser intensity equally 
on all four ground states, which should still yield satisfactory result. In the $\sigma^+\sigma^+\sigma^+\sigma^-$
configuration, we observe a similar arrangement of detunings near the upper F=2 and F=1 states with minor deviations. 
Additionally, $\delta_2$ is switched from a blue detuning to a red one. The cooling force of Scheme 2 is more potent until 2 m/s 
and has a critical velocity of 3.8 m/s. However, we find that Scheme 2 has a peak trapping acceleration that is 64\% of that in Scheme 1.
During our search process, we noticed that even using a small laser intensity for the $\rm L_2$ and $\rm L_3$ laser components, a strong cooling and trapping
force can still be achieved. This observation motivated us to explore a search with only two laser components, namely $\rm L_1$ and $\rm L_4$, 
along with free detunings and laser intensity ratio. Scheme 3 represents the optimal result we discovered. It's different from 
the $\Lambda$-enhanced cooling \cite{cite10} where equal detunings and laser intensity are used for $\rm L_1$ and $\rm L_4$ laser. 
Additionally, we verified that $\Lambda$-enhanced cooling scheme is unable to provide a trapping force. 
Scheme 3 exhibits the strongest cooling force within the range of 2 m/s and has a critical velocity of 3.1 m/s.
Additionally, we observe that the trapping acceleration slope near the MOT center tends to be zero for all the optimal schemes we have identified.

Finally, we make a 3D Monte-Carlo simulation \cite{cite01} with the calculated MOT force for a total of $10^4$ molecules
with a uniform velocity distribution within 1 m/s and a uniform distribution of molecules within a radius of 3
mm. Scheme 1 yields a MOT temperature of \SI{90}{\micro K} and a density of \SI{4.5e8}{cm^{-3}}, while Scheme 2 produces a
temperature of \SI{66}{\micro K} and a density of \SI{4.6e8}{cm^{-3}}. Scheme 3 achieves the lowest temperature of \SI{14}{\micro K} with a density of \SI{4.5e8}{cm^{-3}}.
By reducing the laser power and increasing the field gradient, it should be possible to further improve the system and achieve a temperature 
as low as several \SI{}{\micro K} \cite{cite14} while
increasing the density even further. This would provide an ideal starting point for transferring a large number of
molecules to a conservative trap.

In conclusion, we have presented a general approach for designing a CaF moleculer MOT using Bayesian optimization. 
Our study has yielded three blue-detuned MOT configurations capable of achieving sub-Doppler
cooling within 5 m/s and trapping within 3 mm. Monte-Carlo simulations demonstrate that our MOT schemes
can achieve a temperature as low as \SI{14}{\micro K} and a density as high as \SI{4.5e8}{cm^{-3}}. Our methodology is not limited 
to CaF molecules but can be adapted to other laser-cooled molecules with varying energy level structures to
improve their molecular MOT design.

\section*{Acknowledgements}
%\begin{acknowledgments}
This work was supported by the Innovation Program for Quantum Science and Technology (Grant Nos. 2021ZD0300500 and 2021ZD0300503).
Y. Xia greatfully acknowledge the finacial support from the National Natural Science Foundation of China under Grant Nos. 11834003 and
91836103.
M. Siercke and S. Ospelkaus gratefully acknowledge financial support through  Germany’s Excellence Strategy – EXC-2123/1 QuantumFrontiers. 
%\end{acknowledgments}

%\begin{filecontents}{jobname.bib}

%\end{filecontents}
%\bibliography{jobname}

\bibliography{main.bib}

\end{document}